\documentstyle[osa,manuscript]{revtex}

\begin{document}

\title{
\rightline{{\tt March 1997; Revised March 1998}} 
\rightline{{\tt UM-P-97/09}}
\rightline{{\tt RCHEP 97/02}}
Maximal mixing neutrino models}

\author{J. P. Bowes and R. R. Volkas}
\address{School of Physics\\
Research Centre for High Energy Physics\\
The University of Melbourne\\
Parkville 3052 Australia.}
\maketitle
\begin{abstract}
We account for the solar and atmospheric neutrino problems by introducing 
maximal mixing between conventional neutrinos and sterile 
neutrino partners. We achieve this by invoking a seesaw-like mechanism 
which not only provides us with maximally mixed 
neutrino/sterile-neutrino mass eigenstates but also accounts for 
the relative suppression of the neutrino masses compared to the 
charged fermion masses. In obtaining such an extended seesaw mechanism 
we are required to introduce a new $U(1)$ global symmetry 
together with an extended Higgs sector. 

\end{abstract}

\section{Introduction and Motivation}

The physics of neutrinos continues to be very exciting. Four experiments
(GALLEX, SAGE, Kamiokande and Homestake) have observed a deficit in the
solar neutrino flux \cite{gallex}. Preliminary results from SuperKamiokande
are consistent with the Kamiokande measurements, further strengthening the
case for the existence of a solar neutrino problem \cite{SK}. The most
likely explanation of these results lies in electron-neutrino oscillations
into other species.  Substantial evidence also exists for an atmospheric
neutrino problem \cite{atmos}. In particular Kamiokande and Soudan II see
clear evidence of an anomalously low ratio of muon-like events to
electron-like events. SuperKamiokande data have confirmed this
anomaly, and provided significantly stronger evidence for a zenith-angle
dependence in the atmospheric muon-neutrino flux consistent with an
oscillation based explanation \cite{SK2}. The LSND Collaboration has reported
evidence for both $\overline{\nu}_{\mu} - \overline{\nu}_e$ and $\nu_{\mu} -
\nu_e$ oscillations \cite{lsnd}. This result needs to be checked by an
independent experiment, such as the upgraded KARMEN \cite{kleinfeller}. 

The purpose of the present paper is to construct sensible models which
explain all of the neutrino results using three active and three sterile
neutrino flavours. In order to motivate the introduction of sterile
neutrinos, we first briefly review the valiant attempts that have been made
to explain all of the results using oscillations amongst the three known
neutrino flavours: $\nu_e$, $\nu_{\mu}$ and $\nu_{\tau}$. The Cardall and
Fuller scheme \cite{cardall} sees the atmospheric and LSND results accounted
for through a single $\delta m^2$ of about $0.3\ eV^2$, with the solar
results related to another $\delta m^2$ of about $10^{-5}\ eV^2$. The
atmospheric neutrino anomaly is handled through large-angle $\nu_{\mu} -
\nu_{\tau}$ oscillations. However, the large mass splitting required renders
the atmospheric $\nu_{\mu}$ flux zenith-angle independent. This is
disfavoured by the preliminary Super-Kamiokande data.  The Acker and Pakvasa
three-flavour scheme \cite{acker}, by constrast, accounts for the
atmospheric anomaly through large angle $\nu_{\mu} - \nu_e$ oscillations. 
This scenario is disfavoured by the recent CHOOZ reactor-based bounds
\cite{chooz} (the Palo Verde reactor experiment will check this result
\cite{palo}). It is also disfavoured by the zenith-angle flux dependences
revealed by the preliminary SuperKamiokande data \cite{SK2,fvye}.

Another interesting three-flavour scenario is the maximal mixing model of
Harrison, Perkins and Scott \cite{harrison}. This theoretically appealing
scheme solves the solar neutrino problem through an energy-independent flux
reduction by a factor of $5/9$. The atmospheric neutrino problem is also solved
provided an appropriate $\delta m^2$ is used (for an update see Ref.\cite{fvye}).
However, this model has difficulty accounting for the LSND measurements.

We will assume for the purposes of this paper that all three neutrino
anomalies (solar, atmospheric and LSND) are real. Given this assumption, the
discussion of the previous two paragraphs leads to the conclusion that the
three known neutrino flavours are not sufficient to explain the totality of
neutrino data. This in turn motivates the introduction of sterile neutrinos.

Degree of freedom minimality would be achieved by the introduction of only
one sterile neutrino flavour $\nu_s$. Two different scenarios
result: (i) \cite{sterile}
the solar neutrino problem can be solved through MSW \cite{msw} enhanced
small angle
$\nu_e - \nu_s$ oscillations with $\delta m^2_{es} \simeq 10^{-5}\ eV^2$. 
The atmospheric problem is solved by large-angle or maximal $\nu_{\mu} -
\nu_{\tau}$ oscillations with $\delta m^2_{\mu\tau} \simeq 10^{-3} -
10^{-2}\ eV^2$. The LSND result can be incorporated by small angle $\nu_e -
\nu_{\mu}$ oscillations with $\delta m^2_{e\mu} \simeq 1\ eV^2$. (ii)
Alternatively, the atmospheric problem could be solved by large angle or
maximal $\nu_{\mu} - \nu_s$ oscillations with $\delta m^2_{\mu s} \simeq
10^{-3} - 10^{-2}\ eV^2$, while the solar problem and LSND are handled
through three-flavour $\nu_e - \nu_{\mu} - \nu_{\tau}$ mixing with an
appropriate mass and mixing angle pattern \cite{babu}. 

While the introduction of only one sterile flavour is minimal in regard to
degrees of freedom, it probably is not the most theoretically elegant
possibility. One would have to explain why there is a mismatch between
active and sterile flavours. In addition, for scenario (i) above one would
need to explain why $\nu_e$ and $\nu_{\mu}$ have a small mixing angle
whereas $\nu_{\mu}$ and $\nu_{\tau}$ have a large or maximal mixing angle. 
For scenario (ii) one would want to explain why the sterile flavour mixes
primarily with $\nu_{\mu}$ rather than with $\nu_e$ or $\nu_{\tau}$, or in a
comparable way with all three active flavours.

These issues of theoretical elegance motivate the introduction of three
sterile flavours. We will in addition suppose that the three active flavours
are mixed amongst themselves through small angles. This is motivated by both
the observed pattern of small mixing in the analogous quark sector and the
LSND result. The atmospheric neutrino anomaly then requires that one of the
sterile flavours has a large or maximal mixing angle with $\nu_{\mu}$.  A
theoretical desire to maintain similarity between the generations then
motivates that each of the active flavours is paired with a sterile flavour
through large angle or maximal mixing. 

The neutrino sector of the Exact Parity Model contains parity or mirror
partners, $\nu'_e$, $\nu'_{\mu}$ and $\nu'_{\tau}$, for each of the ordinary
neutrinos \cite{epm}.  Since parity eigenstates must also be Hamiltonian
eigenstates,
maximal mixing between $\nu_{\alpha}$ and its mirror partner $\nu'_{\alpha}$
($\alpha = e, \mu, \tau$) is induced in the limit of small
inter-generational mixing. The Exact Parity Model is therefore a concrete
realisation of the general scenario described in the previous paragraph. The
solar neutrino problem can be solved in a number of ways. (a) The first, simplest and
in many ways most attractive way is via an energy independent $50\%$ flux
reduction through maximal $\nu_e - \nu'_e$ oscillations in the large
parameter range,
\begin{equation}
10^{-10} \stackrel{<}{\sim} \delta m^2_{ee'}/eV^2 \stackrel{<}{\sim} 0.9
\times 10^{-3},
\label{solar}
\end{equation}
where the upper limit is the CHOOZ $\nu_e$ disappearance bound. 
(b) Alternatively, vacuum ``just-so'' oscillations, which require essentially maximal
mixing, can provide an energy-dependent flux
reduction that fits the solar neutrino data well if $\delta m^2_{ee'}$ is fine-tuned
to the $10^{-11} - 10^{-10}\ eV^2$ regime \cite{justso}. (However, we disfavour
this possibility
because of the fine-tuning problem.) (c) It is also possible to take the
inter-generational mixing parameters into the MSW range to achieve a good fit to the
solar neutrino data \cite{wong}. 

The
atmospheric neutrino problem is analogously solved through maximal
$\nu_{\mu} - \nu'_{\mu}$ oscillations in the parameter range
\begin{equation}
10^{-3} \stackrel{<}{\sim} \delta m^2_{\mu\mu'}/eV^2
\stackrel{<}{\sim} 10^{-2}.
\label{atmosp}
\end{equation}
(See Refs.\cite{yasuda} for more precise discussions of the allowed range.)

For solar neutrino scenarios (a) and (b) above, the LSND
result is simply incorporated by switching on small angle mixing between
$\nu_e$ and $\nu_{\mu}$ with $\delta m^2_{e\mu} \simeq 0.1 - 10\ eV^2$. It is at
present not clear how consistent scenario (c) above is with the LSND data.

The purpose of the present paper is to construct analogous models using
strictly sterile rather than mirror neutrinos. In addition to mirror
neutrinos, the Exact Parity Model features parity partners for every
standard fermion, gauge boson and Higgs boson. The mirror sector particles
interact amongst themselves through mirror images of the usual strong and
electroweak interactions. The ordinary and mirror sectors interact through
neutrino mixing (if neutrinos have mass), photon--mirror-photon,
$Z$--mirror-$Z$ and Higgs-boson--mirror-Higgs-boson mixing as well as
gravitationally. The purpose of this paper is to construct a theory with
essentially the same neutrino sector as the Exact Parity Model but without
any of the other pieces of new physics. This is interesting to do for two
main reasons. First, the discussion above has demonstrated how one arrives
at essentially the Exact Parity Model neutrino sector through a
combination of phenomenological and theoretical motivations without ever
invoking parity symmetry as an explicit motivation. Second, it is
interesting to see how elegant or inelegant a model will be which produces
such a neutrino sector in the absence of exact parity symmetry. If, in a few
years time, experiments unequivocally establish the existence of large angle
mixing between active and sterile flavours, then the Exact Parity Model will
be in competition with the models to be constructed herein. In Sec.2 we will
review why maximally mixed active and sterile neutrino pairs are a natural
possibility in the Standard Model with right-handed neutrinos.

Before constructing the models, two remaining issues must be briefly
addressed. First, solving the solar neutrino problem via an
energy-independent $50\%$ flux reduction is considered heterodox. The main
reason for this is that a comparison of Homestake with the Gallium and
(Super-)Kamiokande results
suggests that the mid-energy solar neutrinos, particularly the Beryllium
neutrinos, should be suppressed by more than the high energy Boron and
low-energy $pp$ neutrinos \cite{fits}. A debate has ensued in the literature
about whether or not an energy-independent suppression provides a reasonable
fit to the data.
Note that in addition to the present models, and the Exact Parity Model, the
now disfavoured Acker-Pakvasa scheme and the Harrison, Perkins and
Scott model also feature an energy independent suppression. A proper
treatment of this important issue would require a lengthy discussion. For
the purposes of this paper, we refer the reader instead to the extant
literature \cite{energyindep}. The most important points are the following:
(i) Experiments themselves
will clarify the magnitude of the Beryllium neutrino flux. At present, the greater
suppression claimed for
Beryllium neutrinos depends crucially on the Homestake results.
To be sure that the extra suppression
of mid-energy neutrinos is real, independent experiments must confirm the
Homestake result. Fortunately, the forthcoming Iodine experiment
\cite{iodine} and Borexino \cite{borexino} will do this. (ii) Standard solar
modellers do
not agree on the Boron neutrino flux value, largely due to differing
treatments of the $p + ^7 Be \to ^8 B + \gamma$ cross-section \cite{ssm}. 
It would
help greatly if this theoretical uncertainty could be reduced. In the
interim, our view is that any model which
reduces the solar neutrino flux by a factor like $1/2$ or $5/9$, relative to the
no-oscillation expectation, is worth
serious consideration. (iii) If the present experimental indications in favour of an
energy-dependent flux reduction are confirmed, then the ``just-so'' possibility
discussed earlier might remain viable. Also, the introduction of inter-generational
MSW transitions in addition to averaged maximal $\nu_e - \nu_s$ oscillations would
remain an interesting alternative \cite{wong}.  

Finally, the potential Big Bang Nucleosynthesis (BBN) problem should be
addressed. Suppose both the solar and atmospheric neutrino problems are
solved by maximally mixed active and sterile neutrino pairs.  Then the
sterile state mixing with $\nu_{\mu}$ can be brought into thermal
equilibrium prior to the BBN epoch since the associated $\delta m^2$ is
easily large enough \cite{BBN}. The sterile state mixing with $\nu_e$
destroys successful nucleosynthesis
if the relevant $\delta m^2$ is larger than
about $10^{-8}$ eV$^2$. Current BBN data can just tolerate four equilibrated
relativistic species during the BBN epoch \cite{olive}. So, if the
electronic $\delta m^2$ is less than $10^{-8}$ eV$^2$ there is at present no
problem. Otherwise, there is potentially a problem because the effective
number of neutrinos would be greater than four. However, recent work has
demonstrated that in
a certain region of parameter space active-sterile neutrino oscillations in
the early universe inevitably produce substantial neutrino chemical
potentials or asymmetries \cite{20}. This in turn suppresses active-sterile
transitions via the matter-induced effective potential. Reference \cite{21}
shows that a substantial region of parameter space exists where neutrino
asymmetry creation inevitably suppresses the production of the sterile
states that maximally mix with $\nu_e$ and $\nu_{\mu}$ with $\delta m^2$
values motivated by the two neutrino problems.  This scenario is thus
cosmologically consistent due to a rather subtle but appealing piece of
physics. 

The remainder of this paper is structured as follows. In Sec.2 we review why
maximally mixed active-sterile neutrino pairs are a natural possibility
within the Standard Model augmented by right-handed neutrinos. In Sec.3 
we develop mass matrices that produce both
maximally mixed  and light neutrinos. In Sec.4 we construct in detail
a model which yields one of the suitable mass matrices, and we
comment also on other possibilities. We conclude in Sec.5.

\section{Maximally mixed active and sterile neutrino pairs in the Standard
Model with right-handed neutrinos}

It is interesting to note that maximal mixing between active and sterile
flavours is actually
a natural possibility within the Standard Model augmented by
gauge-singlet right-handed neutrinos \cite{ginnti}.
Consider the usual see-saw mass matrix \cite{seesaw},
\begin{equation}
M_{\nu} = \left( \begin{array}{cc}
0\ & m\ \\ m\ & M \end{array} \right),
\end{equation}
where $m$ is the neutrino Dirac mass and $M$ is the bare
Majorana mass for the right-handed neutrino. In the see-saw model the
limit $m \ll M$ is considered for exemplary reasons. However, the
opposite limit $m \gg M$ is also interesting \cite{ginnti}. In this limit
the mass eigenvalues are 
\begin{equation}
m_{1,2} \simeq m + {M \over 2},\ \ m - {M \over 2},
\end{equation}
and the exact mixing angle is given by
\begin{equation}
\tan 2 \theta = {2m \over M}.
\end{equation}
The $m \gg M$  limit therefore produces
mass eigenstates that are approximately maximal mixtures of 
the active $\nu_L$ and the sterile $(\nu_R)^c$. 
Furthermore, the near degeneracy of the eigenvalues is correlated with
near maximality for the mixing angle ($\theta \simeq \pi/4$) in this
limit, which is an attractive feature given that both the solar and
atmospheric neutrino problems suggest oscillations between nearly
degenerate neutrinos. 

So, one could claim that the goal of this paper is simply achieved in the
above manner. However, the simple scenario above is not completely
satisfactory as a theory of neutrino mass.  As well as the hierarchy $m \gg
M$ we also need $m$ to be tiny compared to the other Dirac masses in the
Standard Model in order to reproduce the correspondingly tiny neutrino
masses required. A satisfactory theory of neutrino mass should explain why
neutrinos are much lighter than all other fermions. The standard see-saw
mechanism can obviously not be used. The models constructed in this paper
will incorporate an extended see-saw mechanism for producing maximally mixed
active and sterile neutrinos {\it that are both also naturally very light.}

\section{Mass Matrices}

The first task in finding maximally 
mixed neutrino/sterile-neutrino models is to search for suitable mass 
matrices. As in the usual 
see-saw mechanism \cite{seesaw} these matrices will include both Dirac 
and Majorana mass terms, and upon diagonalisation will lead to 
Majorana 
mass eigenstates. To find acceptable maximally mixed neutrino/sterile-neutrino 
models we must therefore embark upon a more or less
mathematical exercise in that we wish to find matrices which upon 
diagonalisation satisfy the following three 
requirements:\\\\i) they must have eigenvectors which exhibit maximal 
mixing 
between the neutrino and the sterile neutrino;\\ \\
 ii) the neutrino must have 
a light see-saw suppressed eigenvalue, as in the usual see-saw mechanism; and 
finally, \\\\
iii) the eigenvalues of the neutrino and the sterile neutrino must not 
be equal as we require non-degenerate masses to get oscillations.\\\\

Note that we will concentrate on the single generation case for simplicity. 
In order to solve the solar and atmospheric neutrino problems simultaneously
we of course need at least two generations of maximally mixed active and
sterile neutrinos with $\delta m^2$ values given by Eqs.\ref{solar} and
\ref{atmosp}. We also need intergenerational mixing to be small in order to
not perturb the active-sterile maximal mixing scenario too much. Recall that
small intergenerational mixing is also suggested by the observed form of the
Kobayashi-Maskawa matrix in the quark sector and by the LSND experiment. So,
one can solve all three neutrino anomalies by replicating the one-generation
neutrino sector we will construct below and introducing small
intergenerational mixing. 

In order to get see-saw suppressed eigenvalues, we must in 
addition to our 
sterile neutrino, introduce at least 
one exotic heavy neutral fermion into our model.
In fact the simplest mass matrix which fulfils 
the above properties is the following
$3\times 3$ matrix in $\{\nu_L,(\nu_R)^c,(N_R)^c\}$ space, shown 
below together with its diagonalisation:
\begin{equation}
\label{three}
\Lambda \left(
\begin{array}{ccc}
0&0&a \\
0&0&a \\
a&a&1 
\end{array}\right)
\rightarrow \Lambda\left(
\begin{array}{ccc}
0&0&0 \\
0&-2a^2&0 \\
0&0&1+2a^2 
\end{array}\right),
\end{equation}
where the diagonalising unitary matrix $U$ has the form,
\begin{equation}
U=\frac{1}{\sqrt{2}}\left(
\begin{array}{ccc}
1&1&\sqrt{2}a \\
-1&1&\sqrt{2}a \\
0&-2a&\sqrt{2} 
\end{array}\right).
\end{equation}
Note that we have denoted our 
sterile neutrino as $\nu_R$ and our massive exotic 
neutrino (which is also sterile) as $N_R$, and in the interests of simplicity 
have expressed all masses in terms of $\Lambda$, the mass of 
the heavy exotic neutrino; 
i.e. $a=m_a/\Lambda$.
Thus $a$ is very small and we are hence able to 
express the eigenvalues and eigenvectors in series expanded 
form to order $a^3$ and $a^2$ respectively. 
Although not necessary we have for simplicity also
 assumed that our masses are real. The mass eigenstates 
 $\{\nu_1,\nu_2,\nu_3\}$, corresponding to the 
 mass matrix shown in Eq(\ref{three})
are given by $\{\nu_1,\nu_2,\nu_3\}=\{\nu_L,(\nu_R)^c,(N_R)^c\}U$.

The above $3\times 3$ matrix is however unsuitable for the 
purposes of model 
building as it requires that two of the off-diagonal 
entries appearing in the 
mass matrix, labelled $a$ in Eq(\ref{three}), be equal. 
This is essentially a requirement that the mass terms $\bar{N_R}\nu_L$ 
and $\bar{N_R}(\nu_R)^c$ be equal, which is only possible if we have an 
exact unbroken left-right symmetry. Such a left-right 
symmetric theory 
would however be at odds with the standard model and hence phenomenologically 
ruled out. 
We must hence turn our attention towards suitable $4\times 4$ matrices.

The $2\times 2$ null matrix appearing in the top left hand corner of 
the above $3\times 3$ matrix turns out to be a fundamental requirement 
for any matrix which is to provide us with the required properties. 
It is essentially a generalisation to higher dimension of 
the null entry  appearing in the  
$2\times 2$ matrix
in $\{\nu_L,(N_R)^c\}$ space  
featuring in the usual see-saw mechanism,
\begin{equation}
\left(
\begin{array}{cc}
0&m^{\dagger} \\
m&M  
\end{array}\right).
\end{equation}
Any plausible $4\times 4$ mass matrix 
is therefore expected to be a particular case of 
the following general matrix form 
in $\{\nu_L,(\nu_R)^c,N_L,(N_R)^c\}$ space,
\begin{equation}
\label{gen}
\left(
\begin{array}{cccc}
0&0&m_a&m_b \\
0&0&m_c&m_d \\
m_a&m_c&\Lambda_1&\Lambda \\
m_b&m_d&\Lambda&\Lambda_2 
\end{array}\right).
\end{equation}
In this case we have introduced two heavy sterile
exotic neutral fermions 
$N_L$ and $N_R$ with
 both Dirac $\Lambda$, and Majorana $\Lambda_1$ 
and $\Lambda_2$, mass terms. For simplicity we will again 
express each of the masses 
 in terms of the heavy Dirac mass $\Lambda$ by writing 
$a,b,c,d=m_{a,b,c,d}/\Lambda$, $\lambda_1=\Lambda_1/\Lambda$, and 
$\lambda_2=\Lambda_2/\Lambda$. This 
notation is useful because as we will soon see our 
Dirac mass term $\Lambda$ is constrained to be much 
heavier than all of the other 
mass terms thus allowing us to series expand 
in terms of $a,b,c,d,\lambda_1$, and $\lambda_2$.
 Our task of finding a suitable maximally mixed neutrino model 
is thus essentially 
a task of finding suitable constraints on, or symmetries between, the 
 masses $m_a,m_b,m_c$ and $m_d$.

Four possibilities were considered. The first 
is the simple case where 
all four masses are equal $m_a=m_b=m_c=m_d$. 
However like the $3\times 3$ model previously considered 
this requires that there be 
an unbroken left-right symmetry which is undesirable 
from a model building 
point of view.
Another possibility is to impose a symmetry whereby $m_a=m_d$ 
and $m_b=m_c$; a 
possibility which has already
 been investigated in the context of mirror models \cite{epm}.

The two mass matrices which we will be interested in
 require no symmetry between the mass 
terms. Instead they require the constraint that either
 $m_a=m_d=0$ or $m_b=m_c=0$, with the other 
two parameters being free subject to the condition that 
they be much less 
than $\Lambda$. We thus we need only explain the absence of two of the four 
masses, $m_a,m_b,m_c$ and $m_d$, appearing in Eq(\ref{gen}), hence  
these matrices are desirable from a model building perspective. 
Expanding in terms of the small parameters $a,b,c,d,\lambda_1,$ and 
$\lambda_2$, these two matrices give the following 
diagonalisations, 
\begin{eqnarray}
\label{dir}
&&\left(
\begin{array}{cccc}
0&0&0&b \\
0&0&c&0 \\
0&c&\lambda_1&1 \\
b&0&1&\lambda_2  
\end{array}\right)
\rightarrow \nonumber \\
&&\left(
\begin{array}{cccc}
bc+\frac{1}{2}(b^2\lambda_1+c^2\lambda_2)&0&0&0 \\
0&-bc+\frac{1}{2}(b^2\lambda_1+c^2\lambda_2)&0&0 \\
0&0&1+\frac{1}{2}(\lambda_1+\lambda_2)&0  \\
0&0&0&-1+\frac{1}{2}(\lambda_1+\lambda_2)
\end{array}\right)
\end{eqnarray}
where in this case the diagonalising unitary matrix $U$ takes the form,
\begin{equation}
U=\frac{1}{\sqrt{2}}\left(
\begin{array}{cccc}
-1-\frac{1}{4}\left(\frac{b}{c}\lambda_1-\frac{c}{b}\lambda_2 \right)&
-1+\frac{1}{4}\left(\frac{b}{c}\lambda_1-\frac{c}{b}\lambda_2 \right)&b&-b \\
1-\frac{1}{4}\left(\frac{b}{c}\lambda_1-\frac{c}{b}\lambda_2 \right)&
-1-\frac{1}{4}\left(\frac{b}{c}\lambda_1-\frac{c}{b}\lambda_2 \right)&c&c \\
b&b&1+\frac{1}{4}(\lambda_1-\lambda_2)&-1+\frac{1}{4}(\lambda_1-\lambda_2) \\
-c&c&1-\frac{1}{4}(\lambda_1-\lambda_2)&1+\frac{1}{4}(\lambda_1-\lambda_2)
\end{array}\right),
\end{equation}
and,
\begin{eqnarray}
\label{maj}
&&\left(
\begin{array}{cccc}
0&0&a&0 \\
0&0&0&d \\
a&0&\lambda_1&1 \\
0&d&1&\lambda_2  
\end{array}\right)
\rightarrow  \nonumber \\
&&\left(
\begin{array}{cccc}
ad+\frac{1}{2}(a^2\lambda_1+d^2\lambda_2)&0&0&0 \\
0&-ad+\frac{1}{2}(a^2\lambda_1+d^2\lambda_2)&0&0 \\
0&0&1+\frac{1}{2}(\lambda_1+\lambda_2)&0  \\
0&0&0&-1+\frac{1}{2}(\lambda_1+\lambda_2)
\end{array}\right)
\end{eqnarray}
where our diagonalising unitary matrix $U$ looks like,
\begin{equation}
U=\frac{1}{\sqrt{2}}\left(
\begin{array}{cccc}
-1+\frac{1}{4}\left(\frac{d}{a}\lambda_1-\frac{a}{d}\lambda_2 \right)&
1+\frac{1}{4}\left(\frac{d}{a}\lambda_1-\frac{a}{d}\lambda_2 \right)&a&-a \\
1+\frac{1}{4}\left(\frac{d}{a}\lambda_1-\frac{a}{d}\lambda_2 \right)&
1-\frac{1}{4}\left(\frac{d}{a}\lambda_1-\frac{a}{d}\lambda_2 \right)&d&d \\
-a&-a&1+\frac{1}{4}(\lambda_1-\lambda_2)&1-\frac{1}{4}(\lambda_1-\lambda_2) \\
d&-d&1-\frac{1}{4}(\lambda_1-\lambda_2)&-1-\frac{1}{4}(\lambda_1-\lambda_2)
\end{array}\right).
\end{equation}
Note that in the above expressions we have expanded all 
but the first two entries 
in the diagonalised matrices 
to second order in the small parameters; the first two enteries 
have been expanded to higher order, owing to their increased 
phenomenological interest.
The maximally mixed mass eigenstates 
$\{\nu_1,\nu_2,\nu_3,\nu_4\}$ arising from the above matrices are 
in these cases given by,
\begin{equation}
\{\nu_1,\nu_2,\nu_3,\nu_4\}=\{\nu_L,(\nu_R)^c,N_L,(N_R)^c\}U.
\end{equation}

 From the above diagonalisation it can be seen that we will obtain our 
desired eigenvectors and eigenvalues if we have the  
mass hierarchy $\lambda_1,\lambda_2,a,b,c,d<<1$, 
and the condition that the ratios 
$\lambda_1 b/c,\lambda_1 d/a, \lambda_2 c/b$, and $\lambda_2 
a/d$ be much less than unity. 
Note that the $\Lambda_1$ and $\Lambda_2$ Majorana mass terms are
required to break the degeneracy which would otherwise exist 
between the light eigenvalues.

In the next section we will show one way in which the standard model can be  
extended to account for and explain the mass matrices 
shown in Eq(\ref{dir}) and (\ref{maj}). 
These two matrices are very much analogous, thus  
in the interests of simplicity we will initially concentrate on 
developing a model which gives us a mass matrix like that shown in 
Eq(\ref{dir}). Models which will produce the mass matrix shown in 
Eq(\ref{maj}) will follow trivially.

\section{The Model}

To build an appropriate model we must first of all be able to explain 
the absence of the usual Dirac and Majorana 
mass terms one normally expects to be associated 
with the neutrino $\nu_L$ and its 
sterile partner $\nu_R$, i.e. we must explain the $2\times 2$ null matrix.
The nonexistence of these mass terms can be explained if the Higgs scalars 
required to generate these masses are absent. 

In addition to explaining the absence of these conventional mass terms, we 
must also explain the absence of two of the four possible couplings 
between 
$N_{R,L}$ and $\nu_{R,L}$.
Both of these problems can be solved by introducing a  
new $U(1)$ gauge group with a quantum number which we will label $T$. 
This new $U(1)_T$ symmetry will not only allow us to explain the absence of 
$\nu_L-\nu_R$ mass couplings, but will also allow us to differentiate between 
the mass terms ($m_a\bar{l_L}(N_L)^c$, $m_d\bar{N_R}(\nu_R)^c$) 
and ($m_b \bar{l}_LN_R$,
$m_c\bar{N_L}\nu_R$), as
each class of mass term will transform differently 
under a $U(1)_T$ 
gauge transformation.

The desired neutral mass matrix can be obtained if we assign the following   
$SU(3)_c\otimes SU(2)_L\otimes U(1)_Y\otimes U(1)_T$ group representations 
to our neutral fermions,
\begin{eqnarray}
l_L&\sim &(1,2,-1,0) \nonumber \\
\nu_R&\sim &(1,1,0,2) \nonumber \\
N_L&\sim &(1,1,0,1) \nonumber \\
N_R& \sim &(1,1,0,1).
\end{eqnarray}

The complete list of possible Dirac and Majorana type mass terms which can be 
generated between these neutral fermions are given below:
\begin{eqnarray}
\label{terms}
\bar{l_L}(N_L)^c&\sim &(1,2,1,-1) \nonumber \\
\bar{N_L}\nu_R &\sim & (1,1,0,1) \nonumber \\
\bar{l_L}N_R&\sim &(1,2,1,1) \nonumber \\
\bar{N_R}(\nu_R)^c&\sim &(1,1,0,-3) \nonumber \\
 \nonumber\\
\bar{N_L}N_R&\sim &(1,1,0,0) \nonumber \\
\bar{N_L}(N_L)^c&\sim &(1,1,0,-2) \nonumber \\
\bar{N_R}(N_R)^c&\sim &(1,1,0,-2) \nonumber \\
 \nonumber \\
\bar{l_L}(l_L)^c&\sim &(1,3,2,0) \nonumber \\
\bar{\nu_R}(\nu_R)^c&\sim &(1,1,0,-4) \nonumber \\
\bar{l_L}\nu_R&\sim &(1,2,1,2).
\end{eqnarray}
In our mass matrix we require that the following and only the 
following mass terms 
exist: $m_b\bar{N_R}l_L$, 
$m_c\bar{N_L}\nu_R$, $\Lambda\bar{N_L}N_R$, $\Lambda_1\bar{N_L}(N_L)^c$, 
and $\Lambda_2\bar{N_R}(N_R)^c$.

We can generate these terms to the exclusion of the other terms listed in 
Eq(\ref{terms}) by proposing the existence of the 
Higgs system, 
\begin{eqnarray}
\label{higgs1}
\phi_1& \sim & (1,1,0,1)  \nonumber \\
\phi_2& \sim & (1,2,1,1) \nonumber \\
\phi_3& \sim & (1,1,0,2),
\end{eqnarray}
with vacuum expectation values,
\begin{eqnarray}
\langle\phi_1\rangle&= & v_1  \nonumber \\
\langle\phi_2\rangle&= & \left (
    \begin{array}{c}
    0\\
    v_2
    \end{array}\right ) \nonumber \\
\langle\phi_3\rangle&= & v_3.
\end{eqnarray}
We have made the above assignment in such a way that the Dirac 
mass $\Lambda$ associated with the mass term $\bar{N_L}N_R$ is a bare mass.
This allows for a natural explanation for our required mass hierarchy, 
$\Lambda >>
m_b,m_c,\Lambda_1$,$\Lambda_2$, if the ratios $v_1/\Lambda,v_2/\Lambda,$ 
and $v_3/\Lambda$ are 
much less than unity.

Having demonstrated how one might go about generating the required masses 
for the neutral fermion sector we must now turn our attention to the charged 
fermion sector. 
We wish to generate our charged fermion masses through Yukawa couplings 
involving the $\phi_2 \sim(1,2,1,1)$ Higgs as opposed to the 
usual standard model Higgs 
$\phi\sim(1,2,1,0)$. Our charged fermions are thus given the 
following group assignments,
\begin{eqnarray}
l_L&\sim&(1,2,-1,0) \nonumber \\
e_R&\sim&(1,1,-2,-1) \nonumber \\
q_L&\sim&(3,2,1/3,0) \nonumber \\
u_R&\sim&(3,1,4/3,1) \nonumber \\
d_R&\sim&(3,1,-2/3,-1), 
\end{eqnarray}
and our fermion mass generating Yukawa interactions take the form,
\begin{eqnarray}
\label{yuk}
{\cal L}_{yuk}=g_1\bar{N_L}\phi_1^*\nu_R &+&g_2\bar{l_L}\phi_2^c N_R+
g_3\bar{N_L}\phi_3(N_L)^c+g_4\bar{N_R}\phi_3(N_R)^c \nonumber \\
&+&g_5\bar{l_L}\phi_2 e_R+
g_6\bar{q_L}\phi_2^c u_R+g_7\bar{q_L}\phi_2 d_R+h.c.
\end{eqnarray}
If the vacuum expectation values of $\phi_1,\phi_2$, and
$\phi_3$ are all of similar order, then the smallness of the 
neutrino mass can be explained in a manner analogous to 
that given in  
the standard see-saw mechanism, we retain the usual see-saw relation 
$m_{\nu_{L,R}}m_{N_{L,R}}\simeq m^2_{e,u,d}$.

 From Eq(\ref{higgs1}) it can be seen that the vacuum expectation 
values of $\phi_1$ and 
$\phi_2$ will spontaneously break our $U(1)_T$ symmetry. It is 
thus of interest 
whether our $U(1)_T$ group can be a local symmetry, or whether it 
is confined to be a global symmetry only.
If it is a local symmetry then the Majoron-like Goldstone boson \cite{moh} will be 
eaten by the putative $U(1)_T$ gauge boson. If however it is a 
global symmetry we have to ensure that the Goldstone boson does not 
render our theory phenomenologically unacceptable.
For our gauge group to be locally symmetric it must satisfy each of the five 
$U(1)_T$ anomaly cancellation conditions \cite{anomaly}:
 $[SU(3)]^2U(1)_T$, 
$[SU(2)_L]^2U(1)_T$, $[U(1)_Y]^2U(1)_T$, $U(1)_Y[U(1)_T]^2$, and 
$[U(1)_T]^3$. 
It turns out that there is no way of assigning our new quantum number $T$ 
to our fermion sector in such a way as to satisfy each of these 
anomaly cancellation 
conditions whilst still giving the required neutral fermion mass matrix. 
Our $U(1)_T$ symmetry is hence constrained to be a global symmetry only, 
and we must analyse the imaginary 
components of the Higgs sector to check for any phenomenological difficulties
which may be associated with the Goldstone boson.

Such phenomenological difficulties would arise if it was found 
that our Goldstone boson 
participated in interactions involving 
quarks or charged leptons, or if it coupled to the $Z$ boson.
This would happen if the Goldstone boson field contained an 
admixture of $\phi_2$ in addition to $\phi_1$ and $\phi_3$. Note
from Eq(\ref{yuk}) that $\phi_1$ and $\phi_3$ do not couple to 
charged fermions, and that $\phi_1$ and $\phi_3$ have trivial 
electroweak quantum numbers and hence do not couple to the $Z$ boson.
Since $\phi_1$ and $\phi_3$ have trivial electroweak
quantum numbers, it is fairly clear that the unphysical Goldstone
boson eaten by the $Z$ must be purely from $\phi_2$ and thus the
physical Goldstone boson associated with $U(1)_T$ breakdown will
be constructed purely from $\phi_1$ and $\phi_3$. We now explicitly 
check this reasoning by calculating the neutral mass matrices 
associated with the shifted Higgs fields.
 
We begin by writing the shifted Higgs fields as,
\begin{eqnarray}
\label{17}
\phi_1&= & v_1+\sigma_1+i\sigma'_1  \nonumber \\
\phi_2&= & \left (
    \begin{array}{c}
    0\\
    v_2+\sigma_2+i\sigma'_2
    \end{array}\right ) \nonumber \\
\phi_3&= & v_3+\sigma_3+i\sigma'_3,
\end{eqnarray} 
where the charged Higgs field $\phi_2^+$ has been set to zero
since we know it is eaten by the $W^+$ boson.
Upon substituting Eq(\ref{17}) into the following expression for the 
potential of our three component 
Higgs sector,
\begin{eqnarray}
\label{higgs}
V=\sum_{i=1,2,3}\lambda_{i}(\phi^{\dagger}_i\phi_i)^2
&+&\lambda_4(\phi_1^{\dagger}\phi_1)(\phi_2^{\dagger}\phi_2)+
\lambda_5(\phi_1^{\dagger}\phi_1)(\phi_3^{\dagger}\phi_3)+
\lambda_6(\phi_2^{\dagger}\phi_2)(\phi_3^{\dagger}\phi_3)\nonumber \\
&-&\sum_{i=1,2,3}n_i^2(\phi_i^{\dagger}\phi_i)+ 
a\phi_3^*\phi_1\phi_1+a\phi_3\phi_1^*\phi_1^*,
\end{eqnarray}
and simplifying using the following constraints between  
$v_1,v_2$ and $v_3$, (the minimisation conditions of
 Eq(\ref{higgs})),
\begin{equation}
\label{min1}
\frac{\delta V}{\delta\phi_1}=0=4\lambda_1 v_1^3+2\lambda_4 v_1 v_2^2+
2\lambda_5 v_1v_3^2-2n_1^2v_1+4av_3v_1
\end{equation}
\begin{equation}
\label{min2}
\frac{\delta V}{\delta\phi_2}=0=4\lambda_2 v_2^3+2\lambda_4 v_1^2 v_2+
2\lambda_6 v_2v_3^2-2n_2^2 v_2
\end{equation}
\begin{equation}
\label{min3}
\frac{\delta V}{\delta\phi_3}=0=4\lambda_3 v_3^3+2\lambda_5 v_1^2 v_3+
2\lambda_6 v_2^2 v_3-2n_3^2v_3+2av_1^2,
\end{equation}
we obtain the
following neutral Higgs mass matrices in $(\sigma_1,
\sigma_2,\sigma_3)$ and $(\sigma'_1,\sigma'_2,\sigma'_3)$ space 
respectively,
\begin{equation}
\left(
\begin{array}{ccc}
\label{massi}
4\lambda_1 v_1^2&2\lambda_4 v_1v_2&2\lambda_5v_1v_3+2av_1 \\
2\lambda_4 v_1v_2&4\lambda_2 v_2^2&2\lambda_6 v_2v_3 \\
2\lambda_5v_1v_3+2av_1 &2\lambda_6 v_2v_3&4\lambda_3 v_3^2+av_1^2/v_3 \\ 
\end{array}\right),
\end{equation}
\begin{equation}
\label{massr}
\left(
\begin{array}{ccc}
-4av_3&0&2av_1 \\
0&0&0 \\
2av_1&0&-av_1^2/v_3 \\ 
\end{array}\right).
\end{equation}

Upon diagonalising Eq(\ref{massr}) we find that we are left 
with two zero eigenvalues, one being 
associated purely with the $\sigma'_2$ boson, and the other 
being associated with one of the 
two mixed $\sigma'_1-\sigma'_3$ eigenstates. The remaining orthogonal 
$\sigma'_1-\sigma'_3$ mixed state 
acquires a mass.
The $\sigma'_2$ Goldstone boson 
can thus be gauged away by an $SU(2)_L\times U(1)_Y$ 
local gauge transformation and hence there  
 will be none of the phenomenological 
difficulties otherwise associated with a physical Goldstone boson 
containing an admixture of $\sigma'_2$.
Instead the physical Goldstone boson which inevitably 
arises from the breakdown of the global symmetry $U(1)_T$ 
is a mixed $\sigma'_1-\sigma'_3$ eigenstate 
 involved exclusively in 
 fermion interactions involving at least one of our 
exotic heavy neutral 
fermions $N_L$ and $N_R$. It thus poses no phenomenological problems. 
Likewise the massive CP-odd boson, the orthogonal partner to our 
physical Goldstone boson, 
also presents no difficulties  
as it is both heavy and like the physical Goldstone boson exclusively
associated with interactions involving at least one of our 
heavy neutral fermions. Note that possible cosmological consequences of the
Goldstone boson have yet to be explored.

Finally we briefly consider how the mass matrix in Eq(\ref{maj}) 
can similarly be explained using a $U(1)_T$ extension to the 
standard model. 
To obtain this alternative maximal mixing inducing mass matrix 
we need to include the mass terms $m_a\bar{l_L}(N_L)^c$, and 
$m_d\bar{N_R}(\nu_R)^c$, but exclude 
the $m_c\bar{N_L}\nu_R$, and $m_b\bar{l_L}N_R$ interactions. 
This is 
most easily done by changing the $U(1)_T$ group assignments of our 
exotic neutral fermions $N_L$ and $N_R$ from $T=1$ to $T=-1$. This 
is only a minor alteration to the above model which does not effect 
the Higgs sector or the subsequent analysis in any significant way.

\section{Conclusion}

We have effectively found two independent $4\times 4$ 
mass matrices which upon 
diagonalisation give rise to approximately maximal 
$\nu_L-\nu_R$ mixing \cite{valle}. 
Furthermore, a see-saw-like mechanism is
used in order to explain why two of the four neutrino mass eigenvalues
per generation are tiny. By employing such mass matrices for
the first two generations, the solar and atmospheric neutrino
anomalies can be solved by maximal active-sterile neutrino
oscillations (and the LSND result can be incorporated by introducing
small intergenerational mixing). (As explained in the Introduction,
the potential problem thus created for Big Bang Nucleosynthesis
is elegantly solved by using active-sterile neutrino oscillations
in the early universe to create large neutrino asymmetries \cite{21}.)

The two models developed to theoretically justify these 
two mass matrices both involve the introduction 
of an extended Higgs and  neutral fermion sector, and
a new global gauge group $U(1)_T$. The spontaneous breakdown
of $U(1)_T$ gives rise to a singlet-Majoron-like Goldstone
boson that poses no phenomenological difficulties.

The models presented in this paper should be contrasted with the
Exact Parity Model \cite{epm}. The latter solves the solar and
atmospheric neutrino problems through maximally mixed ordinary
and mirror neutrinos. While the models constructed above have
fewer degrees of freedom than the Exact Parity Model, one has
to conclude that the Exact Parity Model is more elegant and
compelling due to the theoretical appeal of the unbroken
parity symmetry. It is interesting to note that in principle
cosmology could distinguish between the two scenarios. The
neutrino asymmetry creation required to make both the sterile
neutrino and mirror neutrino scenarios consistent with Big
Bang Nucleosynthesis occurs only for certain regions of
parameter space. The parameter space region for mirror
neutrinos is larger than that for truly sterile neutrinos, as
a comparison of the results of Ref. \cite{21} and Ref. \cite{24}
shows. It could turn out that terrestrial experiments 
eventually pinpoint a region that makes mirror neutrinos
cosmologically consistent but truly sterile neutrinos inconsistent.

\section{Acknowledgements}

We thank R. Foot for some useful discussions. RRV is supported by the 
Australian Research Council. JPB is supported by the Commonwealth 
of Australia.

\end{document}